\documentclass[10pt, conference, letterpaper]{IEEEtran}

\IEEEoverridecommandlockouts
\usepackage{cite}
\usepackage{amsmath,amssymb,amsfonts}
\usepackage{graphicx}
\usepackage{textcomp}
\usepackage{xcolor}
\def\BibTeX{{\rm B\kern-.05em{\sc i\kern-.025em b}\kern-.08em
    T\kern-.1667em\lower.7ex\hbox{E}\kern-.125emX}}
\usepackage{algorithm,algpseudocode}
\usepackage{balance,color}
\usepackage{amssymb,amsmath}
\usepackage{graphicx}
\usepackage{epsfig}
\usepackage{grffile}
\usepackage{multirow}
\usepackage{graphicx} 
\usepackage{amsmath}
\usepackage{amsthm}
\usepackage{amsfonts}

\newcommand{\PP}{\mathbb{P}} 
\newcommand{\Fig}[1]{Figure~\ref{fig:#1}}

\DeclareMathOperator*{\argmax}{argmax}
\usepackage{mathtools, nccmath}
\begin{document}

\title{Active Learning-based Classification in Automated Connected Vehicles
}
\author{Alaa Awad Abdellatif$^\dagger$, Carla Fabiana Chiasserini$^{\dagger,\ddagger}$, and Francesco Malandrino$^\ddagger$\\
$^\dagger$ Politecnico di Torino, Italy  -- 
$^\ddagger$ CNR-IEIIT, Italy\\  
}
\maketitle

\begin{abstract}
Machine learning has emerged as a promising paradigm for enabling connected, automated vehicles to autonomously cruise the streets and react to unexpected situations.  A key challenge,  however, is to collect and select  real-time and reliable information for the correct classification of unexpected, and often uncommon, events that may happen on the road. Indeed, the data generated by vehicles, or received from neighboring vehicles, may be affected by errors or have different levels of resolution and freshness. To tackle this challenge, we propose 
an active learning framework that, leveraging the information collected through onboard sensors as well as received from other vehicles,  effectively deals with scarce and noisy  data. In particular, given the available information, our solution selects the data to add to the training set by trading off between two essential  features, namely, quality  and diversity. 
The results, obtained  using real-world data sets, show that the proposed method significantly outperforms  state-of-the-art solutions, providing high classification accuracy at the cost of a limited bandwidth requirement for the data exchange between vehicles. 
\end{abstract}

\begin{IEEEkeywords}
Data selection, connected automated vehicles, online learning.  
\end{IEEEkeywords}

\section{Introduction}\label{Introduction}

The development of automated vehicles and Intelligent Transportation Systems (ITS) has attracted widespread interest in the recent years. Automated vehicles, equipped with cameras, sensors, and lidars, can learn, identify, and handle complex situations occurring on a road, thanks to artificial intelligence (AI) software.  
However, vehicles may face  uncommon situations, such as unexpected maneuvers by neighboring vehicles or movements of pedestrians and bikers, for which limited  history is available. In these cases, conventional AI/supervised learning techniques that rely on large amounts of accurately labeled data for the training, cannot provide  sufficiently  good results. 

\begin{figure}[t!]
	\centering
		\scalebox{1.7}{\includegraphics[width=0.5\columnwidth]{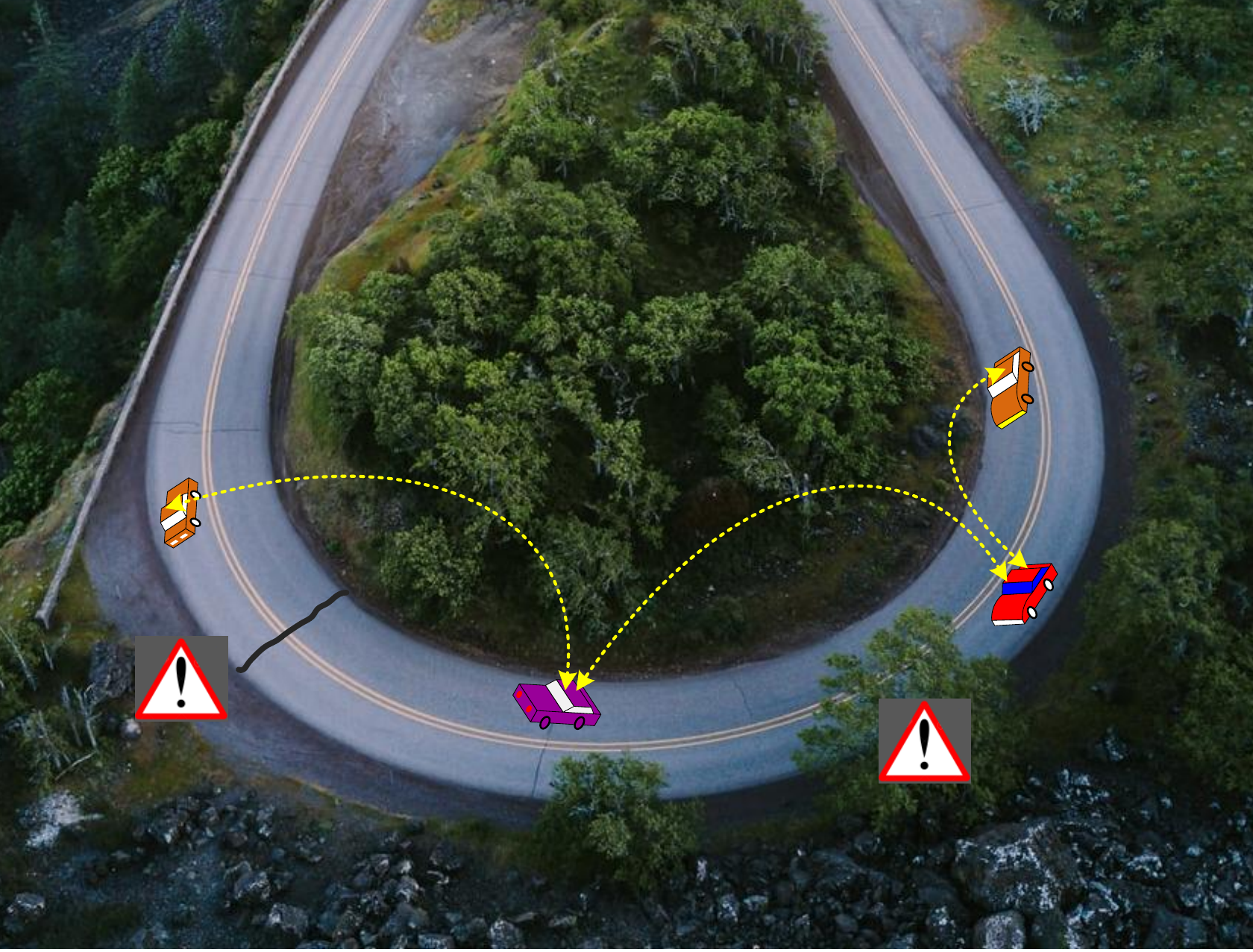}}
	\caption{V2V communications for information exchange about  hazardous/unexpected situations. }
	\label{fig:motivation}
\end{figure}

To overcome such a severe problem, an effective solution consists in (i) exploiting vehicle-to-vehicle (V2V) communication to increase the amount of  data available at a vehicle, and (ii)  adopting on-line machine learning models. Indeed, upon the occurrence of an unexpected event, connected vehicles can, not only leverage the data generated by their own onboard sensors and the history available locally (if any), but also the information received from their neighboring vehicles (see Figure\,\ref{fig:motivation}). 
Additionally, V2V communication can provide a vehicle information on an event well in advance  the vehicle becomes exposed to it, thus allowing for more time to classify the situation and properly react.

As for as on-line learning, Active Learning (AL)  has emerged as a promising technique, which 
actively selects the most informative data to be classified (i.e., labeled) and adds them to the training set \cite{AL2017}. 
Hence, the training set and learning model are updated progressively and iteratively, so as to continuously improve the quality of the classification decisions  \cite{ASPAL2018}.  
Traditionally, AL schemes rely on the presence of a super-accurate classifier that generates the ground-truth for unlabeled data, an  assumption that turns out to be impractical in many real-time applications, such as connected, automated vehicles. Indeed, vehicles are typically weak labelers, and often have at their disposal  noisy data (e.g., generated by cameras in the presence of fog or rain),  or data that may have  different levels of resolution  and freshness. 
Thus, not only the labels generated by the vehicles' classifiers, but also the data generated or received by a vehicle, may be affected by errors.

In this paper, we tackle the above challenges by proposing an AL framework for connected vehicles, which selects a subset of the locally-generated and received data, to be used for classification.
Our solution not only addresses the problem of data scarcity by leveraging V2V communications, but it also carefully identifies the information that can  effectively improve the classification accuracy at each vehicle, accounting for data diversity as well as label and data quality.
More specifically, our main contributions are as follows: 
\begin{enumerate}
	\item we  investigate  three ways in which vehicles can leverage the information exchanged through V2V communications for online training, namely, by sharing labels, data, or a combination of the two. For each of these operational modes, we study the impact on the classification accuracy as well as on the network load;
	\item we propose a method to define the information quality, including two main steps: (i) label integration, to generate an aggregate label for the acquired data, and (ii) data quality assessment, to measure the quality of the acquired data based on labelers' accuracy, data freshness, and affinity of the corresponding labels with respect to the aggregate label;      
	\item we define a data selection scheme, which accounts for the trade-off between data quality and diversity, in order to obtain a maximally diverse set of data with high quality;
	\item we evaluate the proposed framework  and compare it against state-of-the-art solutions, using real-world datasets. Our results demonstrate the effectiveness of the proposed approach in improving the classification performance. 
\end{enumerate}

The rest of the paper is organized as follows. Sec.\,\ref{sec:Related} discusses the related work and highlights the novelty of our study. Sec.\,\ref{sec:System} describes the system model, while Sec.\,\ref{sec:Methodology} presents the proposed AL framework, along with the label integration and the data selection schemes. Sec.\,\ref{sec:Performance} introduces the scenario and the data traces we used for the performance evaluation, and it shows  the obtained results and the gain with respect to existing solutions. Finally,  Sec.\,\ref{sec:Conclusion} concludes the paper.


\section{Related Work}\label{sec:Related}
 
Most of the existing works on AL address the problem of noisy  (or imperfect) labels in binary classification \cite{niosyBinaryClass2016,niosyBinaryClass2015}, while very  few tackle the multi-class (i.e., multi-label) case. Among the latter   ones, \cite{DeepLearningFromCrowds,Imbalanced_Label2015,Subset_selection2019}  investigate the classification performance of crowdsourced data, where labeling can be done by volunteers or non-expert labelers. 
In particular, \cite{DeepLearningFromCrowds} presents an AL-based, deep learning technique, leveraging volunteered geographic information to overcome the lack of  big data sets; therein, a customized loss function is specifically defined to effectively deal with  noisy labels and avoid performance degradation. 
 \cite{Subset_selection2019} enhances the performance of supervised learning with noisy labels in crowdsourcing systems by applying the majority voting label integration method and selecting the data referring to the same event whose label is sufficiently close to the resulting value. 
The work in \cite{Imbalanced_Label2015} studies the problem of   imbalanced noisy labeling, where the available labeled data are not evenly distributed across the different classes. 
First, it performs 
label integration and data selection  based on data uncertainty and class imbalance level, then it classifies unlabeled data  using the trained model and adds them to the training set. 

A different application is instead  targeted in \cite{ASPAL2018}, where AL is used for  incremental face identification. The study therein aims to build a classifier that progressively selects and labels the most informative data, and then it adds the newly labeled data to the training set. Furthermore,  AL is combined with  self-paced learning (SPL) -- a recently developed learning scheme that gradually incorporates from easy to more complex data into the training set, with easy data being those with high classification confidence.  
Note that all the above works consider specific classifiers (or loss functions), which cannot be easily incorporated in other learning techniques \cite{multiclass2018}; thus, finding a  label integration and data selection strategy that can be integrated with a generic  multi-class classification scheme, is still an open problem.

With regard to cooperative applications leveraging V2V communications, it is worth noticing that several car manufacturers have already enabled their vehicles to share real-time hazard signals and  to automatically alert each other \cite{AutonomousVehicles}. The integration of V2V communication with machine learning to improve road safety has also received significant interest. In particular, \cite{PacketLoss} studies the impact of communication loss on 3D object detection exploiting a deep-learning approach. In  \cite{VehicularComm}, Gaussian process regression is used to estimate the age of the vehicles' data and proactively allocate, e.g., transmission power and resource blocks for reliable and low-latency V2V communication.      
\cite{VehicleRecognition2019} deals with the vehicle type recognition problem, where labeling a sufficient amount of data  is very time consuming. 
The solution presented in \cite{VehicleRecognition2019} consists in exploiting fully labeled web data to reduce the labeling time of surveillance data through deep transfer learning; also, only images of unlabeled vehicles' with  high uncertainty and diversity are selected to be queried. 

{\bf Novelty.} Our work is the first to address the problem of scarce and noisy data available to vehicles for classifying unexpected events. Furthermore, unlike other studies, we assess the data quality level accounting  for many factors, including freshness. Finally, the method we propose for label integration and data selection is general enough to be incorpotrated in different classification techniques or loss functions. 


\section{System Model}\label{sec:System}

{\bf Vehicles and topology.}
We denote as~$\mathcal{V}$ the set of all vehicles, and as~$\mathcal{E}(t)$ the set of edges, i.e., pairs of vehicles within radio range of each other. Given an {\em ego vehicle}~$v_0\in\mathcal{V}$, we indicate as~$\mathcal{N}_{v_0}(t)$ the set of neighbors of~$v_0$, i.e., vehicles~$v\in\mathcal{V}\colon (v_0,v)\in\mathcal{E}(t)$. The road topology is divided into discrete {\em segments}~$i\in\mathcal{I}$. Time is continuous, and~$t_{ij}$ denotes the time at which vehicle~$j\in\mathcal{V}$ is found at segment~$i\in\mathcal{I}$.

{\bf The classification task.}
In our scenario, each vehicle has its own active learning process running locally, which combines locally- and remotely-generated information. Such information comes from onboard cameras and ADAS sensors; in the following, we refer to both sensor readings and features extracted from information as {\em data}. The high-level purpose of the adaptive learning system we consider is to {\em classify} such data, associating each of them with a {\em label}. Specifically, $x_{ij}$ denotes the data observed by vehicle~$j\in\mathcal{V}$ while traveling at segment~$i\in\mathcal{I}$, and~$y_{ij}\in\mathcal{L}$ the associated label ($\mathcal{L}$~is the set of all possible labels). Data observed by the ego vehicle and the locally-generated labels thereof are indicated by~$x_{i0}$ and~$y_{i0}$ respectively. The combination of data, label, and time~$(x_{ij},y_{ij},t_{ij})$ is called a {\em sample}.

{\bf Modes.}
As mentioned already, vehicles can receive information from their neighbors. Specifically, we compare the following three {\em modes}:
\begin{itemize}
    \item {\bf labels}: the ego vehicle receives only the set~$\mathcal{Y}_i=\{y_{i0},y_{i1},\dots,y_{iJ}\}$ of labels generated by the vehicles~$\{1,2,\dots,J\}\in\mathcal{N}_{v_0}(t)$;
    \item {\bf data}: the ego vehicle receives only the set~$\mathcal{X}_i=\{x_{i0},x_{i1},\dots,x_{iJ}\}$ of data observed by the vehicles~$\{1,2,\dots,J\}\in\mathcal{N}_{v_0}(t)$;
    \item {\bf samples}: the ego vehicle receives both labels and data.
\end{itemize}
It is important to observe that modes significantly differ in the network usage they imply. Indeed, labels can be orders of magnitude smaller than the data they are based upon; hence, the ``labels'' mode is potentially much more efficient than the ``data'' and ``samples'' ones. 


\begin{figure*}[t!]
	\centering
		\scalebox{3.4}{\includegraphics[width=0.27 \textwidth]{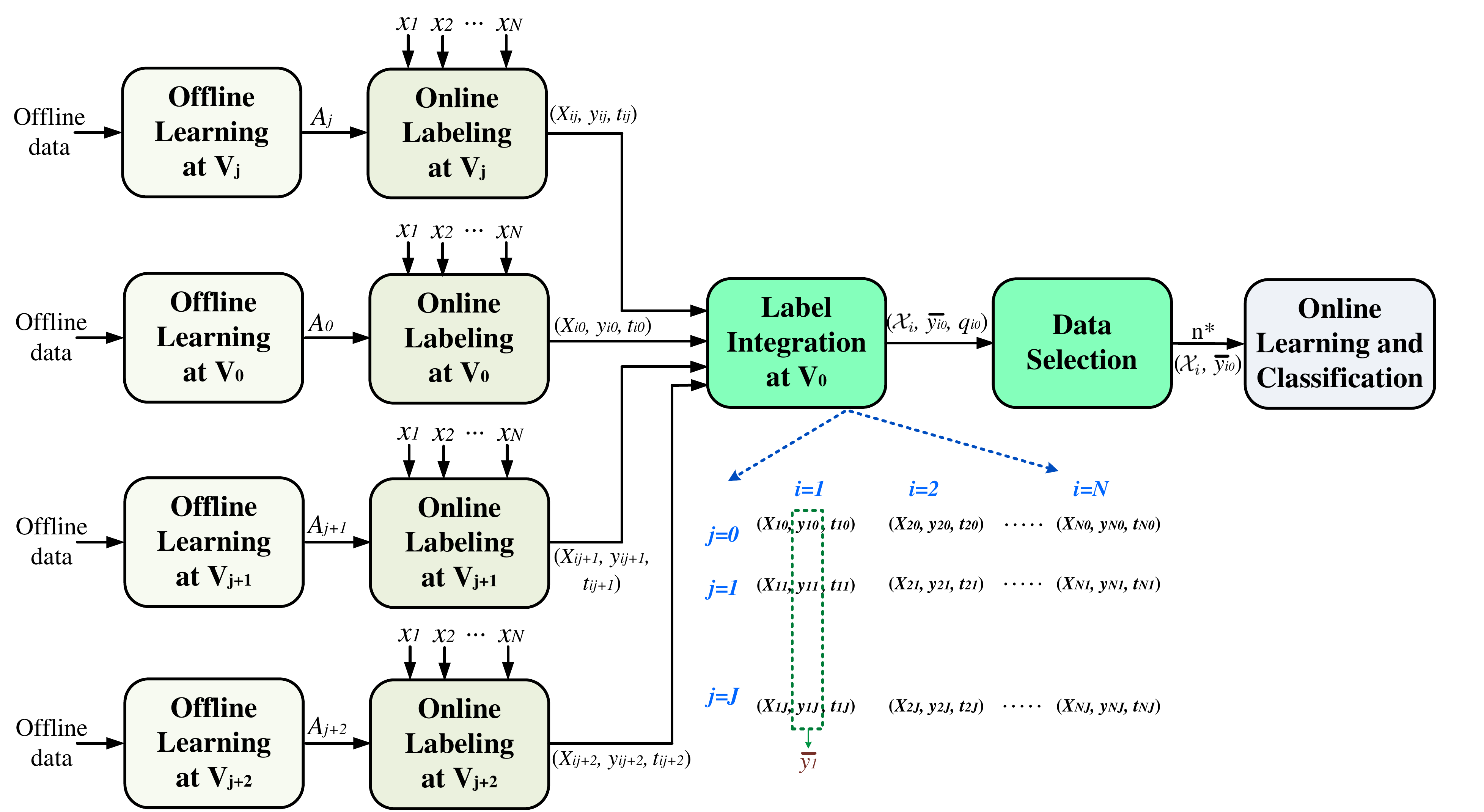}}
	\caption{Diagram representing the proposed AL framework, highlighting the different stages performed by the vehicles and the corresponding data flow }
	\label{fig:flowchart}
\end{figure*}

\section{Active Learning Framework}\label{sec:Methodology} 

In this section, we first  introduce the proposed AL framework and the  performance metrics we consider. Then we detail the two core procedures of our framework: label integration and data selection.

\subsection{Methodology and performance metrics}

Our framework, depicted in Figure \ref{fig:flowchart}, includes three  main stages, as described below.

\textbf{Offline learning:} 
We consider that each vehicle has a certain amount of history, composed of $M$ samples,  through which it can initially train its learning model; as mentioned, $M$ may be very small. With reference to such off-line data set, we define the accuracy  of the generic vehicle labeler $j$ as,
\begin{equation}
A_j = \frac{\sum_{m=1}^{M} I(y_{mj}, {y_m})}{M}.  
\label{eq:accuracy} 
\end{equation}
where $y_{mj}$ is the label generated by $j$ for sample $m$,  $y_m$ is the  estimated label, and $I(u, w)$ is an indicator function, such that $I(u, w)=1$ if $u=w$, and 0 otherwise.  

\textbf{Online labeling:} 
Let us now consider that an event takes place at time $t_{i0}$ while the ego vehicle $v_0$ is in road segment $i$. The vehicle acquires some information  through its onboard sensors and, possibly extracts some features (we  refer to such new information as {\em data}); then $v_0$ labels such data through the learning model,  obtaining  {\em sample} $(x_{i0}, y_{i0}, t_{i0})$. Depending on the adopted operational mode, the ego vehicle shares with its neighbors,  labels, data, or samples. Note that, in the data operational mode, $v_0$ has to labels the received information using its own training model.


\textbf{Label integration:} 
After receiving the information from the neighboring vehicles, $v_0$ computes an aggregated label for the acquired data, using one of the label integration strategies reported below. Clearly, in label and sample mode, $v_0$ leverages the labels received from other vehicles, while in data mode it exploits the labels $v_0$ itself has obtained for  the data locally generated and for the received data.
We denote the aggregated label with $\bar{y_{i0}}$; moreover, we define quality indicator $q_{i0}$, which accounts for the accuracy of the vehicles $A_j$ from which $v_0$ receives information and for the data freshness $f_{ij}$, as detailed in the following subsections.      The samples referring to the situation  in road segment $i$ to which the ego vehicle is exposed are therefore described as $(\mathcal{X}_{i0}, \bar{y_{i0}}, q_{i0})$, where $\mathcal{X}_{i0}$ is the  data referring to such an event available at $v_0$.  

\textbf{Data selection and classification:} 
The ego vehicle selects the most appropriate set of samples to update its learning model. The goal of our data selection scheme is to find a maximally diverse collection of samples (with respect to all possible labels, i.e., data classes) in which each sample has as high quality as possible.     
 Using the selected samples, $v_0$ updates its training model and performs classification.

\subsection{Label integration and quality definition\label{sec:integration} }
We consider and compare the following label integration techniques for the computation of the aggregate label $\bar{y_{i0}}$. 

{\bf Majority Voting (MV).}   
The simplest and most popular label integration method is MV \cite{Imbalanced_Label2015}, which assumes no prior knowledge on the labelers' accuracy or data freshness. In MV,   $\bar{y_{i0}}$ is computed as: 
\begin{equation}  
\bar{y_{i0}}^{(MV)} = \argmax_{l \in \mathcal{L}} \sum_{j=0}^{J} I(y_{ij}=l) \,. 
\label{eq:Label_MV}
\end{equation}
Given $\bar{y_{i0}}^{(MV)}$, the quality indicator of a sample, $q_{i0}^{MV} \in\left[0,1\right]$, is defined as  
\begin{equation}
q_{i0}^{MV} =  \frac{1}{J+1} \left( \max_{l\in \mathcal{L}} \sum_{j=0}^{J} I(y_{ij}=l) \right),    
\label{eq:q_MV}
\end{equation}
Besides neglecting data freshness, MV's performance is acceptable when more than $50\%$ of the labelers have high accuracy, which does not always hold in complex real-world scenarios \cite{Imbalanced_Label2015}. Thus, in what follows, we propose alternative techniques that aim to overcome MV's weaknesses. 

{\bf Weighted Majority Voting (WMV).} 
We now associate a probability of correctness, $p_{ij}$, representing the probability with which $v_0$ receives a correct information from  $v_j$. Such probability depends on the labeler's accuracy and data freshness: 
\begin{equation}
p_{ij} =  f_{ij} \cdot A_j,    
\end{equation}  
with the data freshness being defined as: 
\begin{equation}
f_{ij} = \left\{
                \begin{array}{ll}
                  \exp[-(t_{i0} - t_{ij})]  & t_{i0} > t_{ij} \\
                  0   & t_{i0} \leq t_{ij} 
                \end{array}
              \right.
\end{equation}
with the values of $f_{ij}$ ranging from $0$ (totally stale data) to $1$ (absolutely fresh data) \cite{datafreshness}. 
Considering that $p_{ij}$'s are independent with respect to $j$ \cite{WMV2017}, we write:
\begin{equation}
\bar{y_{i0}}^{(WMV)} = \argmax_{l\in \mathcal{L}} \PP(\bar{y_{i0}}^{(WMV)}=l | y_{i0}, \cdots, y_{iJ}).   
\end{equation}  
Following the standard hypothesis testing procedure \cite{WMV2017} and assuming for ease of presentation binary classification with equal priors,  $\PP(\bar{y_{i0}}^{(WMV)}=1)=\PP(\bar{y_{i0}}^{(WMV)}=-1$, the aggregate label $\bar{y_{i0}}^{(WMV)}$ is given by:
\begin{equation} 
\bar{y_{i0}}^{(WMV)} = \left\{
                \begin{array}{ll}
                  1, & \rho_1  > \rho_2 \\
	         	0 &  \rho_1 = \rho_2 \\
                  -1,  & \rho_1  < \rho_2  
                \end{array}
              \right. 
\label{eq:WMV}
\end{equation} 
where $\rho_1 = \prod_{\{j:y_{ij}=1\}} p_{ij}$, and $\rho_2 = \prod_{\{j:y_{ij}=-1\}} p_{ij} $.  
Hence, the quality indicator of WMV is defined as 
\begin{equation}
q_{i0}^{WMV} =  \max_{l\in \mathcal{L}} \prod_{\{j:y_{ij}=l\}} p_{ij}.  
\end{equation}

{\bf Weighted Average (WA)} 
The WA  method relies on defining a weighting coefficient accounting for both the labelers' accuracy and data freshness: 
$\lambda_{ij} = a \cdot f_{ij} + b \cdot A_j$, where $a$ and $b$ are constants representing the  importance of $f_{ij}$ and $A_j$, respectively.    
Then, the aggregate label is defined as   
\begin{equation}     
\bar{y_{i0}}^{WA} = \argmax_{l\in \mathcal{L}} \sum_{j=0}^{J} \exp(\lambda_{ij}) \cdot I(y_{ij}=l) \,.
\label{eq:WA}
\end{equation} 
We remark that the exponential function in (\ref{eq:WA})  is used to weight more the labels associated with high classification confidence, thereby making it more descriptive than a simple average.    
The quality indicator of WA is then given by: 
\begin{equation}
q_{i0}^{WA} =  \max_{l\in \mathcal{L}} \sum_{j=0}^J \exp(\lambda_{ij}) \cdot I(y_{ij}=l).
\label{eq:QWA}
\end{equation}

To assess the performance of the proposed label integration methods, we define the Labeling Accuracy (LA)  for the ego vehicle as  
\begin{equation}
LA_0 = \frac{\sum_{i=1}^{\Omega} I(\bar{y_{i0}}, {\gamma_i})}{\Omega} 
\end{equation}
where $\Omega$ is the size of the testing data set for the type of event currently occurring in road segment $i$, and $\gamma_i$ is the ground truth.

We remark that WMV and WA methods account for the labelers' quality and freshness; also, WA leverages an exponential function with a weighting coefficient to magnify the effect of high-quality labelers, which improves the performance compared to MV and WMV.

\subsection{Data subset selection}

The objective of our data selection algorithm, named Quality-Diversity Selection (QDS), is to obtain a subset of high-quality data to be added to the online training set so as to maximize the model classification accuracy. We highlight that, unlike most of the existing quality-based schemes to data selection that result in a reduced samples' diversity, our  approach efficiently  trades-off quality and diversity, thus significantly improving the performance of the AL framework.   
Furthermore, the QDS algorithm not only determines {\em which} samples should be selected but also {\em how many}  should be added to the training set. 

Let us first define the average quality score of the selected samples as: 
$Q(x) = \frac{\sum_{i=1}^{n} {q_{i0}}}{n}$,  
where $n$ is a number of selected samples and $q_{i0}$ is sample quality indicator (defined in Sec.\,\ref{sec:integration}). 

Then the diversity score is measured based on the entropy of the selected samples \cite{Entropy}: 
$ H(x) = -\sum _{i=1}^{K} \kappa_{i} \log_2 \kappa_{i} $, 
where $\kappa_{i}$ is the proportion of samples belonging to class $i$, and $K$ is the number of classes. 
Accordingly, the sample selection is  conducted in two steps: 
\begin{itemize}
\item \textbf{Selecting  the class  $k^*$:} the class of samples to target is chosen so as to maximize diversity, i.e., 
$k^* = \argmax_{k} H(\mathcal{X})$. 
The idea is indeed that the more diverse samples are selected, hence the more balanced their distribution across the different classes, the more informative they will be. 
\item \textbf{Selecting the samples:} given  class $k^*$, the samples with the best quality are selected such that the quality score is maximized, i.e., 
$\mathcal{X}^* = \argmax_{\mathcal{X}} Q(\mathcal{X})$. 
\end{itemize}
  
The selected samples are added to the on-line training set and the classification accuracy of the AL framework is checked by labeling the data in the testing set. If the obtained classification accuracy $\hat{\alpha}$ is below the desired predefined value $\alpha$, one more sample is selected, till accuracy $\alpha$ is reached. 
The proposed QDS algorithm is summarized in Algorithm \ref{alg:alg_QDS}, where $n^*$ is the number of selected samples. 

\begin{algorithm}
\caption{Quality-Diversity Selection (QDS) Algorithm}
\label{alg:alg_QDS}
\begin{algorithmic}[1]
\State {\textbf{Input:} $(\mathcal{X}_{i0}, \mathcal{Y}_{i0}, \mathcal{T}_{i}), \forall i \in {1, \cdots N}$. }
\State {Compute $\bar{y_{i0}}$ and $q_{i0}$ using a label integration method (e.g., MV, WMV, WA)}
\State {Identify the selected class $k^*$} 
\State {Given $k^*$, select samples with maximum quality $\mathcal{X}^*$} 
\State {Add selected samples to the online training set $\mathcal{O}$ }
\State {Compute $\hat{\alpha}$}
\If {$\hat{\alpha} > \alpha$}
\State {$n^*=\left|\mathcal{O}\right|$}
\State{Break}  \Comment{$n^*$ is obtained}  
\EndIf
\State \Return{ $n^*$, $\mathcal{O}$}
\end{algorithmic}
\end{algorithm}

\section{Performance Evaluation}\label{sec:Performance}

For our performance evaluation, we use two datasets: the one in~\cite{vehiclesType}, consisting of a set of photos of four types of vehicles (namely, a double decker bus, a Cheverolet van, a Saab 9000, and an Opel Manta 400), and the one in \cite{MNIST}. The images were processed with the BINATTS image processing system, extracting a combination of scale-independent features through a combination of classical moment-based measures and heuristic ones such as hollows, circularity, rectangular and compactness.

In all simulations, we assume that the ego vehicle~$v_0$ is helped by a total, of four labelers. Furthermore, in order to model the fact that vehicles may have different {\em quality} levels (e.g., quality of their sensors, camera, and computational capabilities), each vehicle is assigned a different classification model~\cite{MachineLearning} (one of: fine tree, medium tree, linear SVM, medium Gaussian SVM, and weighted KNN classifier), with the best-performing classifiers associated with the highest quality.

\begin{figure}
\centering
\includegraphics[width=.3\textwidth]{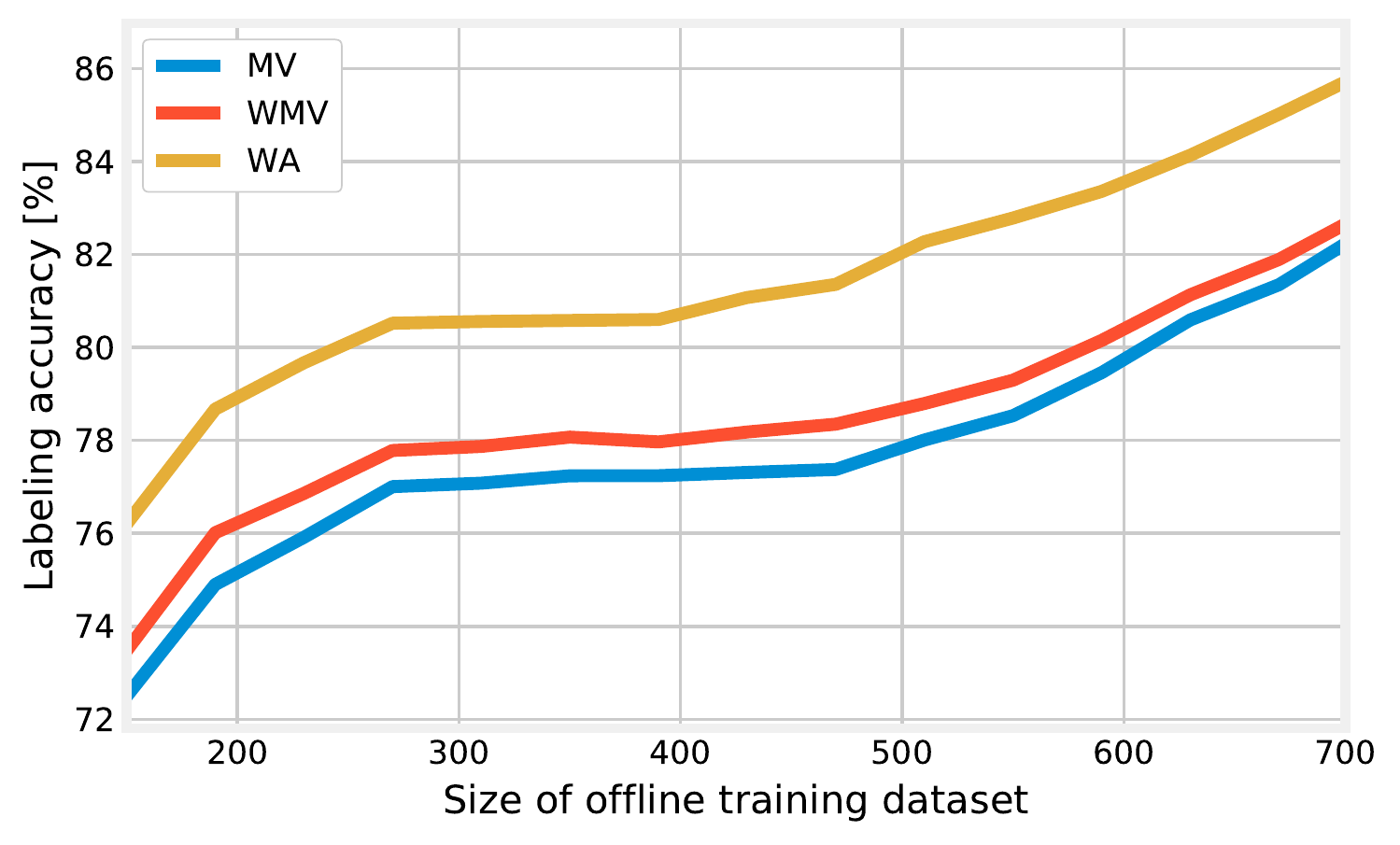}
\caption{Labeling accuracy as a function of the size~$M$ of the offline training set, for different label integration methods.
    \label{fig:labelintegration}
} 
\end{figure}

\begin{figure*}
\centering
\includegraphics[width=.3\textwidth]{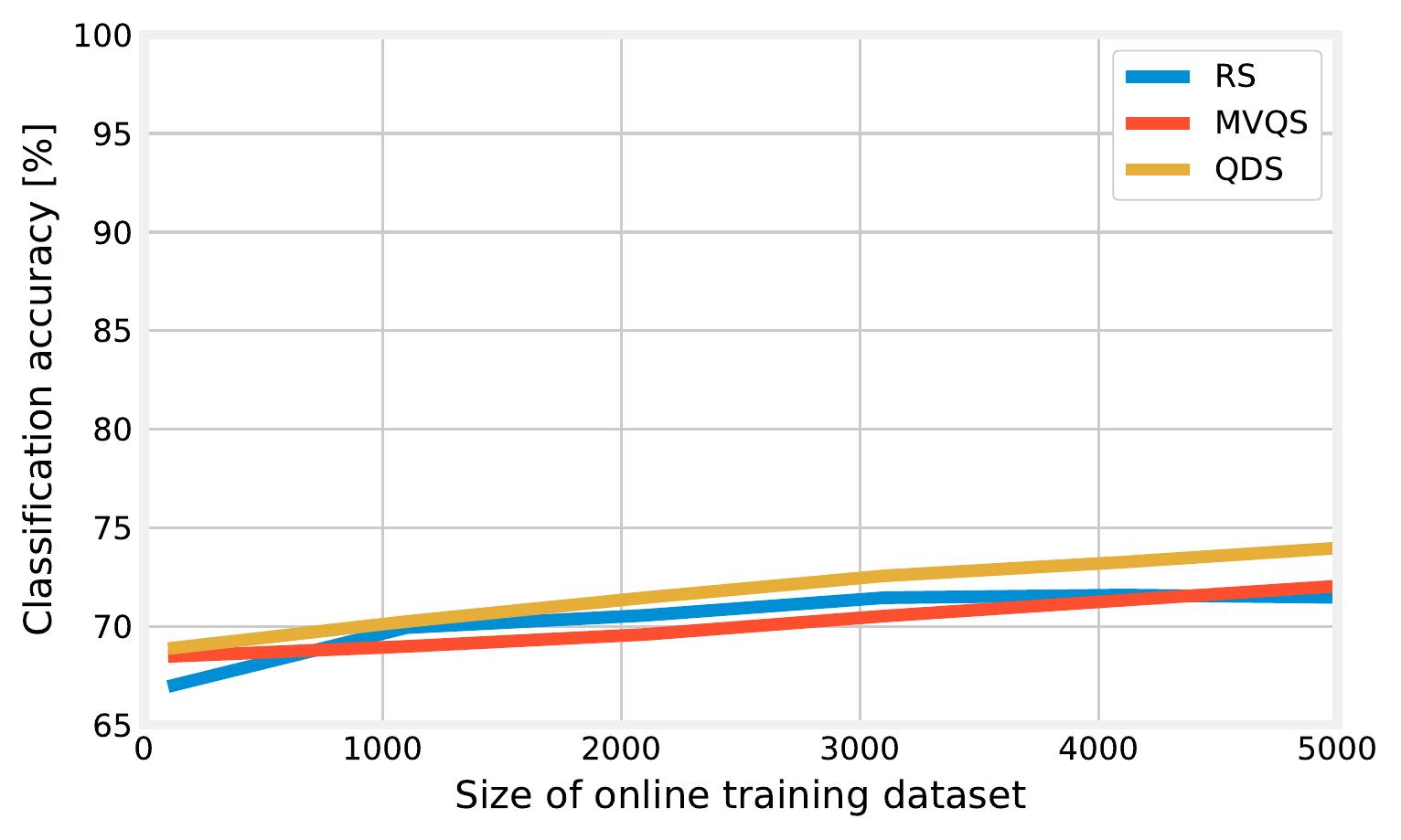}
\includegraphics[width=.3\textwidth]{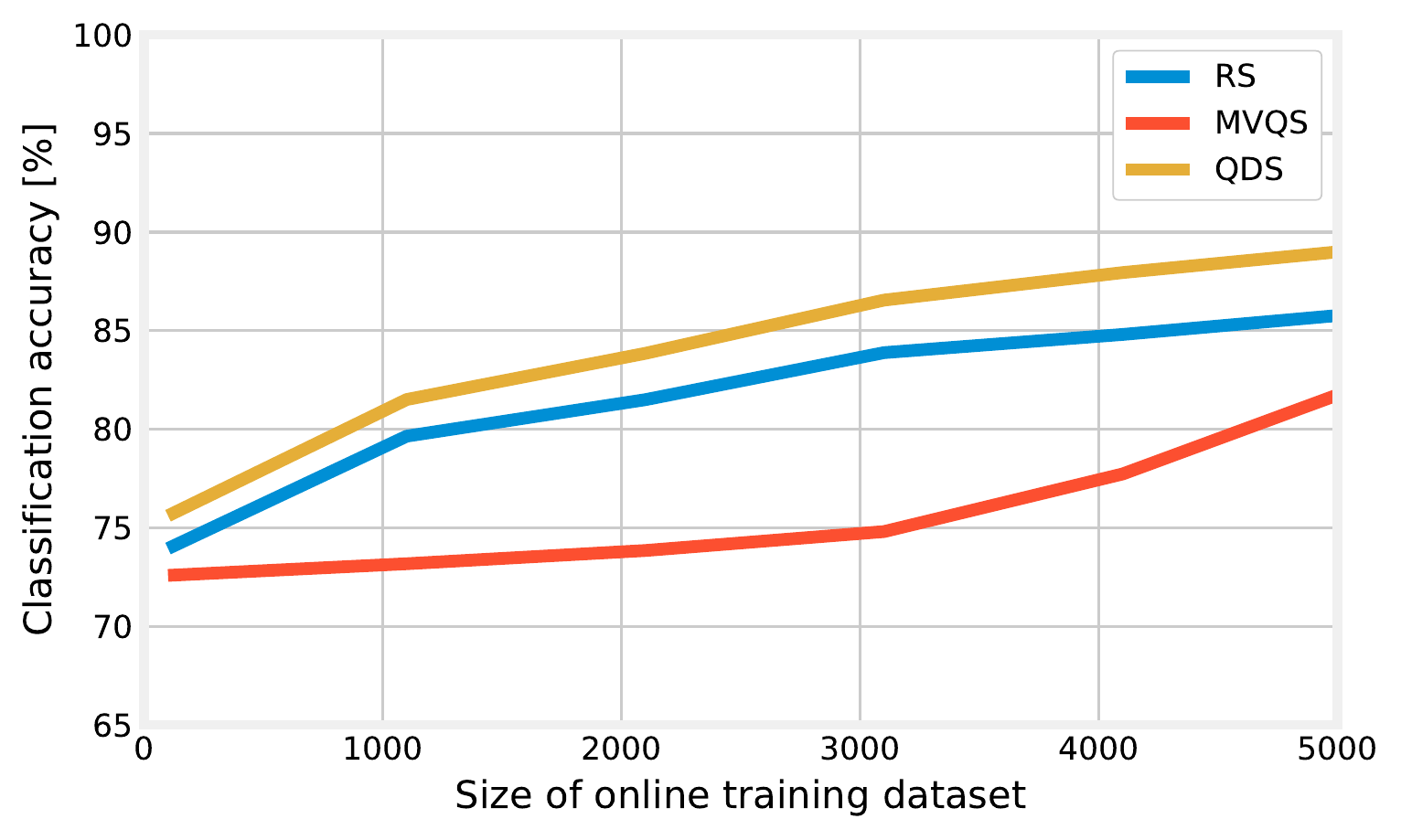}
\includegraphics[width=.3\textwidth]{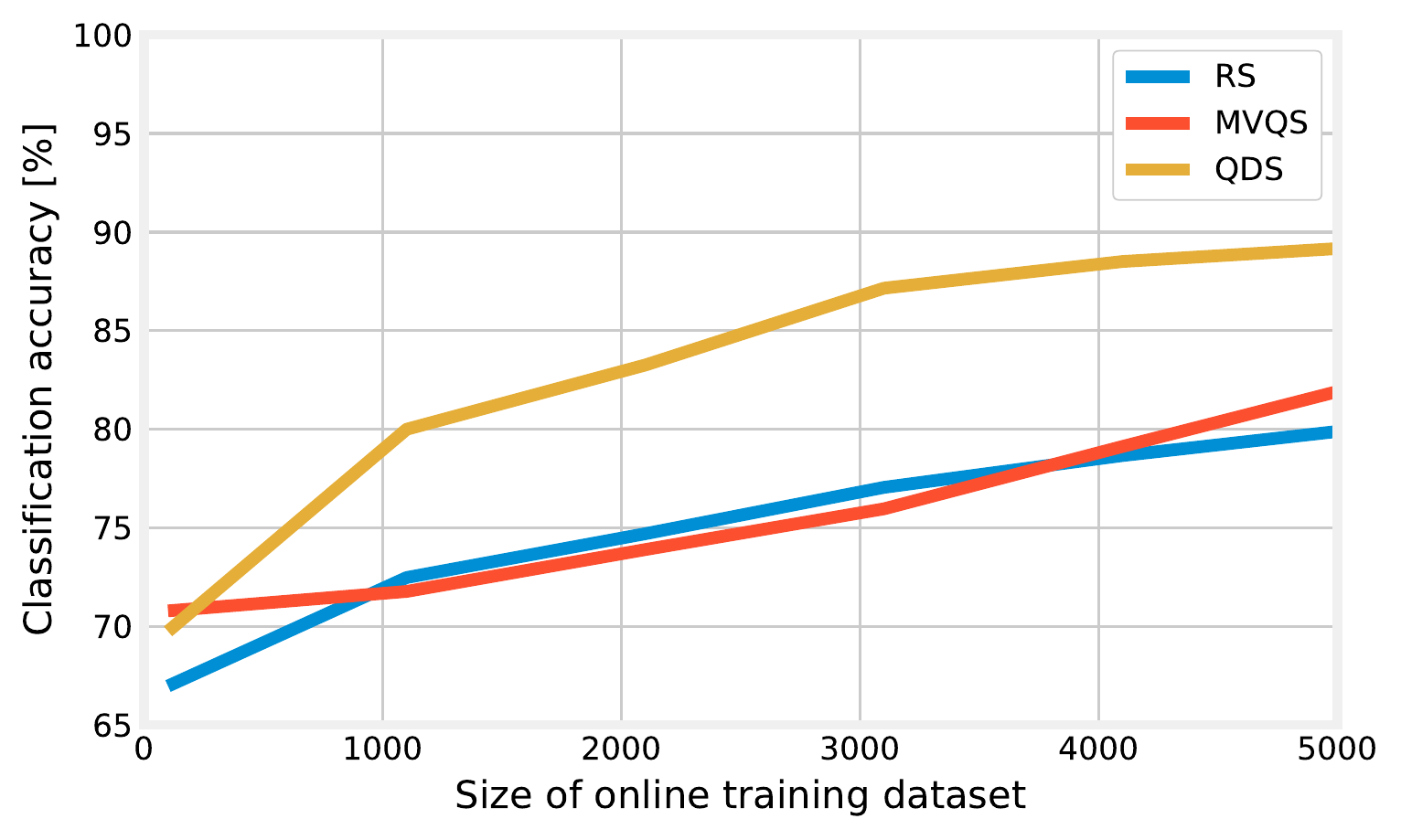}
\caption{
    Classification accuracy as a function of the size of the online training set, for different data selection strategies, under the labels (left), data (center), and samples (right) modes.
    \label{fig:dataselection}
} 
\includegraphics[width=.3\textwidth]{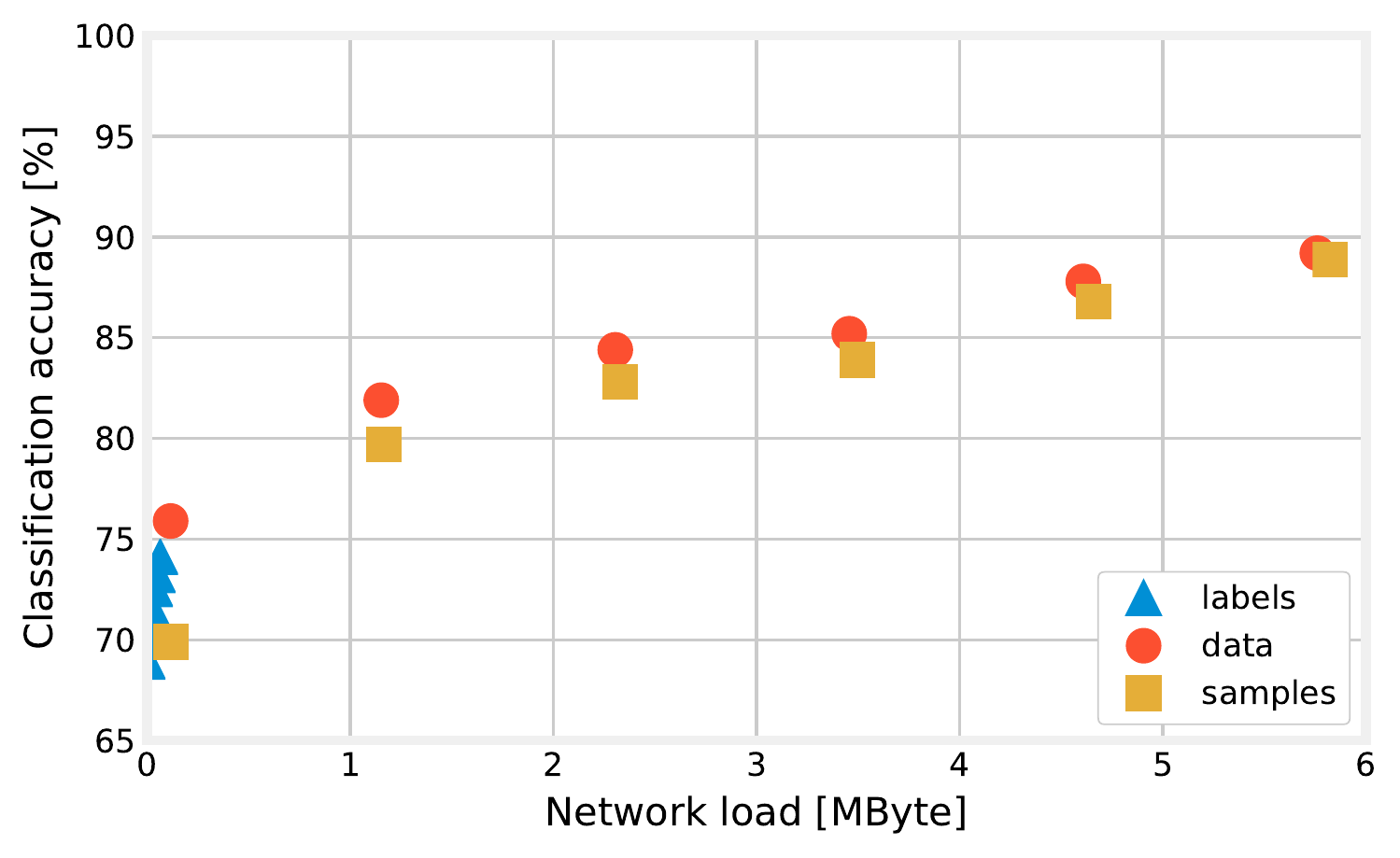}
\includegraphics[width=.3\textwidth]{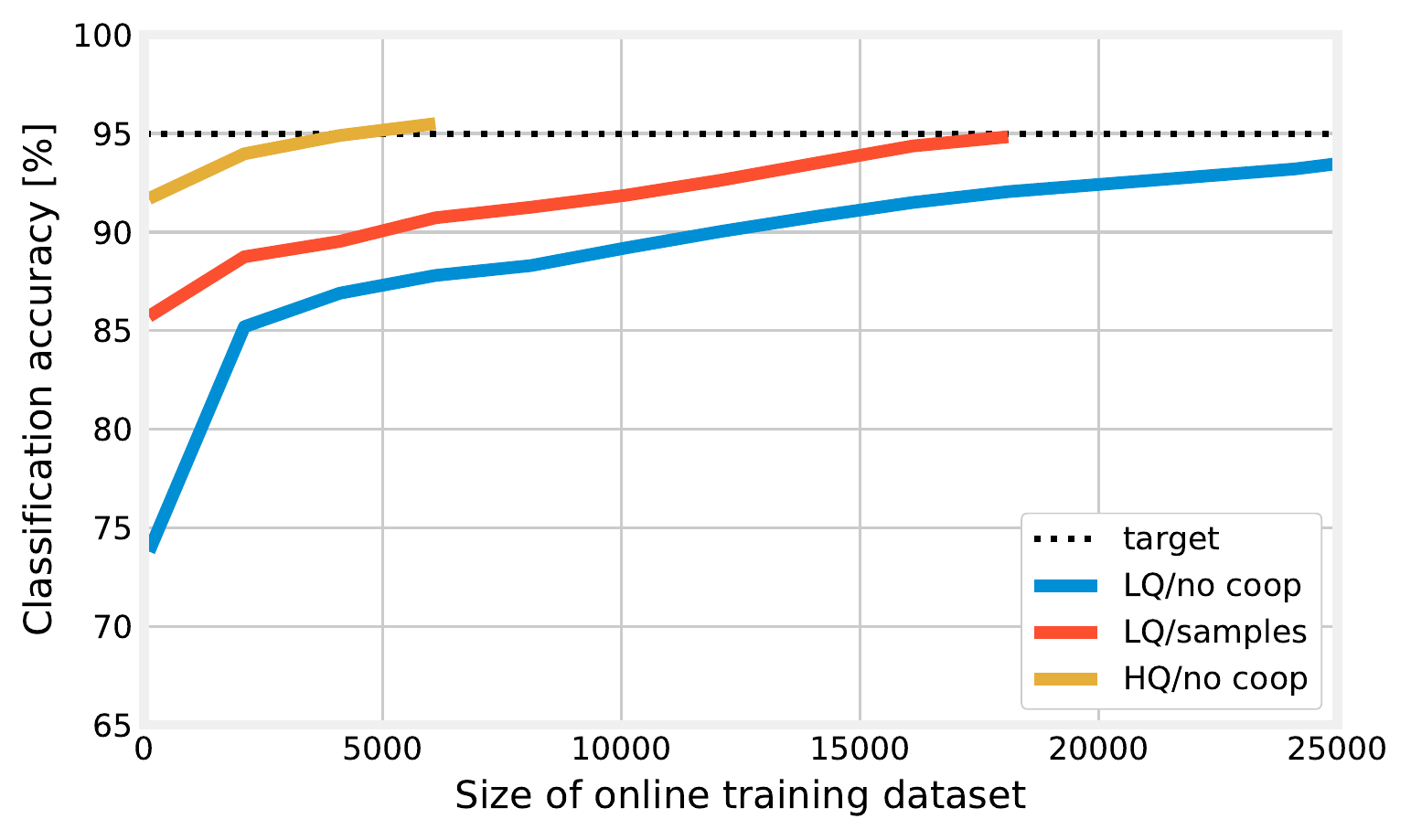}
\includegraphics[width=.3\textwidth]{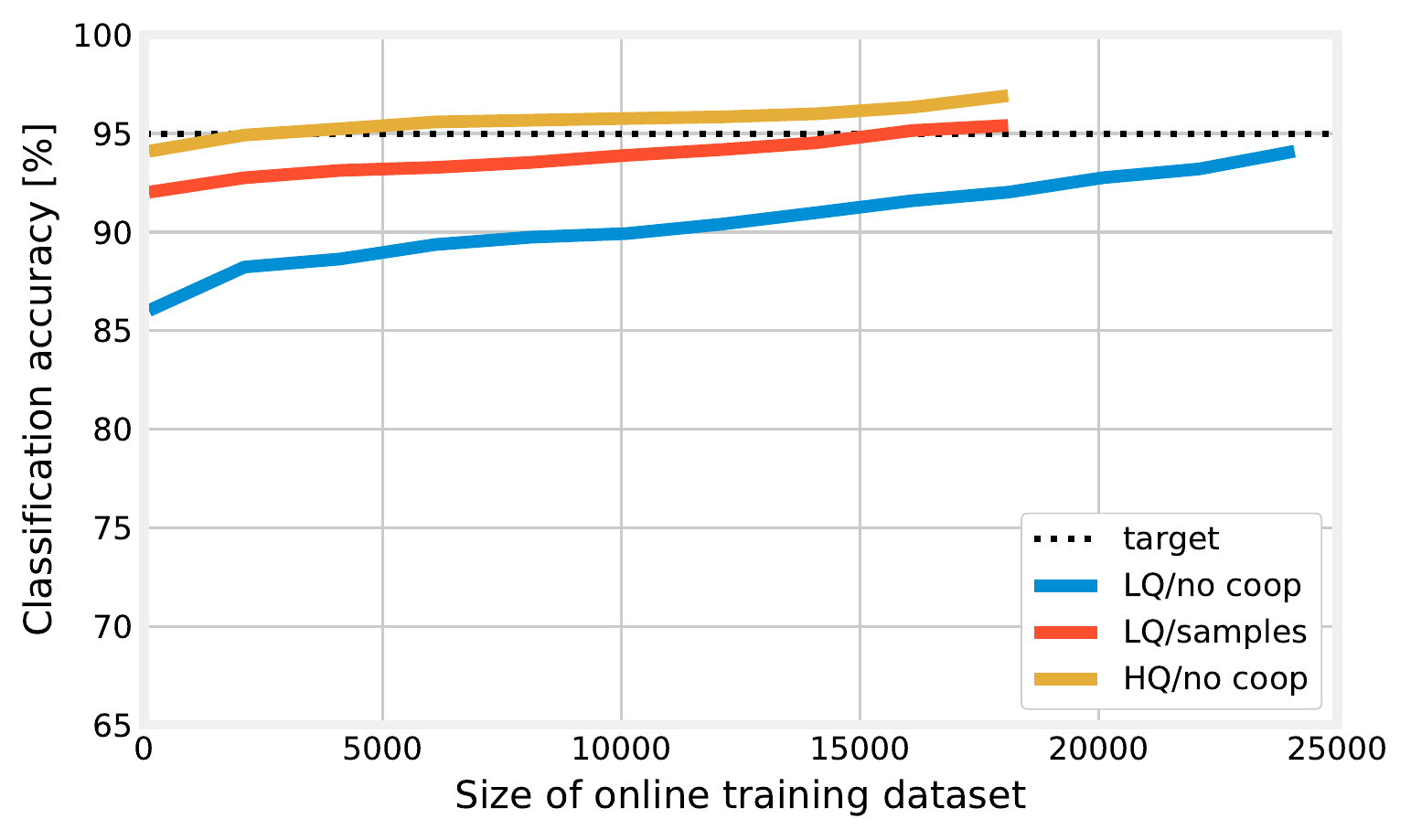}
\caption{
Trade-offs between classification accuracy and network load (left); effect of cooperation on classification accuracy when the size of the offline training set is~$M=500$ (center) and~$M=2000$ (right).
    \label{fig:loadcoop}
} 
\end{figure*}

The first aspect we are interested in is the impact of the label integration method on the labeling accuracy. To this end, \Fig{labelintegration} presents the LA as a function of the size~$M$ of the offline training set, using the {\em vehicle} dataset. It is possible to observe how a larger training set always corresponds to a better accuracy; more interestingly, the weighed average integration (WA) yields substantially better accuracy than majority voting (MV) and weighed majority voting (WMV). An intuitive explanation is that WA is able to use all available samples, while at the same time accounting for their quality and freshness. Based on this result, we use the WA integration method for the remainder of our results.

We now study, in \Fig{dataselection}, how the data selection algorithm and the mode of operation influence performance. The plots show the classification accuracy as a function of the online training set size; each curve therein corresponds to a data selection algorithm, and each plot corresponds to a different mode. Comparing the individual lines within each plot, it is possible to observe how our own QDS algorithm consistently outperforms both the state-of-the-art approach MVQS \cite{Subset_selection2019}, where highest-quality samples are selected for training, and the baseline random-selection approach (RS). This result suggests that sample quality is not the only factor to consider when assembling a training set, and that labeler's accuracy shall be considered as well. 
Looking at the three plots, it is clear that the samples mode is associated with higher performance than the data mode, and both outperform the labels mode; consistently with one's intuition, more information -- be it labels or data -- translates into better performance.

The better performance of the data and samples modes comes, however, at a cost of an increased network load, as summarized \Fig{loadcoop}(left). Each marker therein corresponds to a combination of mode and training size, and its x- and y-coordinates (respectively) correspond to the network load and the achieved classification accuracy. The figure highlights how different trade-offs between network load and classification accuracy can be pursued and that, in general, the two quantities are strongly correlated.

Last, we turn to the issue of cooperation between vehicles, and to assess how much, and for whom, cooperation is beneficial. The plots \Fig{loadcoop}(center)--\Fig{loadcoop}(right) show how, for a fixed size of the offline training set, the quantity of available {\em online} training data influences the classification accuracy; different lines correspond to high-quality (HQ/no coop) and low quality (LQ/no coop) vehicles with no cooperation, as well as to low-quality vehicles operating in samples mode (LQ/samples). Cooperation yields a substantial performance advantage for low-quality vehicles, which can reach the desired level of accuracy ($\alpha=0.95$ in the plots) with a substantially smaller number of samples, hence, in a much shorter time.

\section{Conclusion}\label{sec:Conclusion}

We proposed an active learning framework for connected automated vehicles, which leverages vehicle-to-vehicle communication to increase the amount of collected data to be added to the training set. Given that a vehicle can receive multiple data, labels, or a combination of the two from its neighbors, we proposed  label integration methods and a data selection algorithm, which account for the labelers' accuracy, the data freshness, and the data diversity.  We evaluated our approach using real-world data sets, and we showed that it outperforms state-of-the-art solutions.




\balance
\bibliographystyle{IEEEtran}
\bibliography{alaaref}

\end{document}